\documentclass[preprint,pra,showpacs,nofootinbib]{revtex4}
\usepackage{amssymb}
\usepackage{latexsym}
\usepackage{epsfig}
\usepackage{subfigure}
\usepackage{makecell}
\usepackage{amsmath}
\usepackage[colorlinks=true,linkcolor=red]{hyperref}
\newcommand{\bea}{\begin{eqnarray}}
\newcommand{\eea}{\end{eqnarray}}
\newcommand{\beq}{\begin{equation}}
\newcommand{\eeq}{\end{equation}}

\begin{document}

\title{CMB power spectrum in the emergent universe with k-essence}
\author{Qihong Huang$^{1}$\footnote{Corresponding author: huangqihongzynu@163.com}, Kaituo Zhang$^{2}$, He Huang$^{3}$, Bing Xu$^{4}$ and Feiquan Tu$^{1}$}
\affiliation{
$^1$ School of Physics and Electronic Science, Zunyi Normal University, Zunyi, Guizhou 563006, China\\
$^2$ Department of Physics, Anhui Normal University, Wuhu, Anhui 241000, China\\
$^3$ Institute of Applied Mechanics, Zhejiang University, Hangzhou, Zhejiang 310058, China\\
$^4$ School of Electrical and Electronic Engineering, Anhui Science and Technology University, Bengbu, Anhui 233030, China
}

\begin{abstract}
The emergent universe provides a possible method to avoid the big bang singularity by considering that the universe stems from an stable Einstein static universe rather than the singularity. Since the Einstein static universe exists before inflation, it may leave some relics in the CMB power spectrum. In this paper, we analyze the stability condition for the Einstein static universe in general relativity with k-essence against both the scalar and tensor perturbations. And we find the emergent universe can be successfully realized by constructing a scalar potential and an equation of state parameter. Solving the curved Mukhanov-Sasaki equation, we obtain the analytical approximation for the primordial power spectrum, and then depict the TT-spectrum of the emergent universe. The results show that both the primordial power spectrum and CMB TT-spectrum are suppressed on large scales.
\end{abstract}

\maketitle

\section{Introduction}

While most of problems in the standard big bang cosmological model can be solved by the inflationary scenario~\cite{Guth1981, Linde1982, Albrecht1982}, the big bang singularity problem at the beginning of universe remains open. To avoid the big bang singularity, by considering a form of energy named as \textit{quintessence} which is a dynamic, time-dependent canonical scalar field $\phi$ with a potential $V(\phi)$ to drive the late-time cosmic acceleration, a cosmological model called \textit{emergent universe} was proposed~\cite{Ellis2004a, Ellis2004b}. In the emergent universe, it is assumed that the universe is originated from an Einstein static universe rather than a big bang singularity. After the universe exits from the Einstein static state, it can evolve into an inflationary era and then exit from this era. When the emergent universe was proposed, it had gotten lots of attention~\cite{Campo2007, Wu2010, Cai2012, Zhang2014, HuangQ2015, Shabani2017, Shabani2019, Huang2020}. For a successful emergent universe, it requires the Einstein static universe can exist past-eternally, i.e. it is stable against both scalar perturbations and tensor perturbations. However, the original model of emergent universe is unsuccessful since the Einstein static solution in general relativity with quintessence is unstable against inhomogeneous scalar perturbations~\cite{Barrow2003}. Subsequently, a large amount of efforts have gone into theories of modified gravity, and the stable Einstein static universe against homogeneous and inhomogeneous scalar perturbations was found in Mimetic gravity~\cite{Huang2020}, scalar-fluid theory~\cite{Bohmer2015}, non-minimal derivative coupling model~\cite{Huang2018a, Huang2018b}, braneworld model~\cite{Zhang2016}, Jordan-Brans-Dicke theory~\cite{Huang2014}, Eddington-inspired Born-Infeld theory~\cite{Li2017}, hybrid metric-Palatini gravity~\cite{Bohmer2013}, GUP theory~\cite{Atazadeh2017}, f(R,T) gravity~\cite{Sharif2019}, f(R,T,Q) gravity~\cite{Sharif2018} and massive gravity~\cite{Li2019}.

In slow-roll inflation, nearly scale-invariant primordial scalar perturbations caused by quantum fluctuations during inflation can explain the cosmic microwave background (CMB) radiation anisotropy observed today and provide seeds for the large-scale structure of the observable Universe~\cite{Lewis2000, Bernardeau2002}. CMB observations show that there exists a suppression of CMB TT-spectrum at large scales, which was first observed by COBE~\cite{Smoot1992} and recently confirmed by Planck 2018~\cite{Planck2020}. This might correspond to the physics before inflation. In order to explain this suppression, several approaches are proposed. One approach is to introduce the spatial curvature in the inflationary model~\cite{Bonga2016, Handley2019}. Recently, by considering the universe starting with a kinetically dominated regime followed by a slow-roll epoch, it was found that the suppression of CMB TT-spectrum exists in general relativity by considering the spatial curvature~\cite{Thavanesan2021,Shumaylov2022}. Another approach is to construct some new models, such as, pre-inflation~\cite{Dudas2012, Cai2015}, pre-inflationary bounce~\cite{Cai2018}, non-flat XCDM inflation model~\cite{Ooba2018}, warm inflation~\cite{Arya2018}, Double inflation~\cite{Feng2003}, hybrid new inflation~\cite{Kawasaki2003}, emergent universe~\cite{Labrana2015}, and so on. In the emergent universe, when the Einstein static state is assumed as a superinflating phase, the suppression of CMB TT-spectrum at large scales was realized~\cite{Labrana2015}. In this work, the positive curvature is just used to ensure the existence of Einstein static state rather than having an influence on the primordial perturbations. Taking into consideration both the effect of the positive curvature on the primordial perturbations and the postulate of the universe originating from an Einstein static state followed by a slow-roll inflation, the suppression of CMB TT-specturm at large scales is weakened~\cite{Huang2022}. It is also worthy to be noted that these work in Ref.~\cite{Labrana2015} and ~\cite{Huang2022} are analyzed in general relativity with quintessence, and the Einstein static universe is unstable against the inhomogeneous scalar perturbations. So, the emergent universe fails to explain the suppression of CMB TT-spectrum in general relativity.

\textit{K-essence} is characterized by a scalar field with a non-canonical kinetic term~\cite{Armendariz-Picon2000, Armendariz-Picon2001}. It was first proposed as a model for inflation~\cite{Armendariz-Picon1999, Garriga1999}, and then as a model for dark matter~\cite{Scherrer2004, Bose2009}. After it was proposed, it received lots of attention and was widely studied in cosmology, such as, k-essence cosmology~\cite{Aguirregabiria2004}, classical stability~\cite{Abramo2006}, behavior in phase space~\cite{Yang2011}, slow-roll conditions~\cite{Chiba2009}, thermodynamic properties~\cite{Bilic2008}, and so on. In cosmological observation, it was found that k-essence was hard to be distinguished from quintessence~\cite{Barger2001} but it might have some imprints on the perturbation spectrum~\cite{Malquarti2003}. In addition, when the spatial curvature is considered in k-inflation, it was found that the CMB TT-spectrum is suppressed~\cite{Shumaylov2022}. Thus, in this work, we plan to explore whether a successful emergent universe, in which the universe stems from a stable Einstein static state and then evolves into a slow-roll inflation, can be realized in general relativity with k-essence, and then study whether the suppression of CMB TT-spectrum can be realized in the emergent universe.

The paper is organized as follows. In Section II, we give the field equations and the Einstein static solutions. In Section III, the stability of Einstein static solutions against tensor perturbations is analyzed. In Section IV, we study the stability conditions under the homogeneous and inhomogeneous scalar perturbations. In Section V, we design how the universe exits from the Einstein static state, evolves into a slow-roll inflationary epoch. In Section VI, we solve the curved Mukhanov-Sasaki equation and obtain the analytical primordial power spectra of the emergent universe, and then we plot the CMB TT-spectra of the emergent universe. Finally, our main conclusions are shown in Section VII.

\section{Field Equations and Einstein static solutions}

In this section, we begin with the general action
\beq\label{action}
S=\int d^{4} x \sqrt{-g}\Big[\frac{1}{2}R+P(X,\phi)\Big]+S_{m},
\eeq
with
\bea
X=-\frac{1}{2}g^{\mu\nu} \nabla_{\mu}\phi \nabla_{\nu}\phi, \quad P(X,\phi)=\alpha X-V(\phi),
\eea
where $R$ is the Ricci curvature scalar, $V(\phi)$ is the potential of the scalar field $\phi$, and $S_{m}$ represents the action of a perfect fluid. $\alpha$ is a coupling parameter, $\alpha=1$ and $\alpha=-1$ corresponds to the case of quintessence and phantom, respectively.

Varying the action~(\ref{action}) with respect to $g_{\mu\nu}$ and $\phi$, we obtain the Einstein field equation and the scalar field equation
\bea
&& G_{\mu\nu}+\frac{\alpha}{2}g_{\mu\nu} \nabla^{\alpha}\phi \nabla_{\alpha}\phi+g_{\mu\nu} V-\alpha \nabla_{\mu}\phi \nabla_{\nu}\phi-T_{\mu\nu}=0,\nonumber\\
&& \label{EF}\\
&& \nabla_{\mu}(\alpha g^{\mu\nu} \nabla_{\nu}\phi)-V_{,\phi}=0,\label{SF}
\eea
where $V_{,\phi}=\frac{dV}{d\phi}$, $T_{\mu\nu}$ is the energy-momentum tensor of the perfect fluid.

To find an Einstein static solution, we consider a closed Friedmann-Lemaitre-Robertson-Walker(FLRW) universe
\bea
&& ds^2=-a(\eta)^2d\eta^{2}+a(\eta)^2\gamma_{ij}dx^i dx^j,\nonumber\\
&& \gamma_{ij}dx^i dx^j=\frac{dr^2}{1-r^2}+r^2 (d\theta^2+sin^2 \theta d\varphi^2),
\eea
where $a$ represents the scale factor and $\eta$ denotes the conformal time. Then, the $(00)$ and $(ij)$ components of Eq.~(\ref{EF}) give
\bea
&& \mathcal{H}^{2}+1=\frac{1}{3}\Big(\frac{\alpha}{2}\phi'^{2}+a^{2} V+a^{2} \rho\Big),\label{00}\\
&& 2\mathcal{H}'+\mathcal{H}^{2}+1=-\Big(\frac{\alpha}{2} \phi'^{2}-a^{2} V+a^{2} p\Big),\label{ij}
\eea
where $'=d/d\eta$, $\mathcal{H}=a'/a$, $\rho$ and $p$ are the energy density and the pressure of the perfect fluid which satisfies $p=\omega\rho$ with $\omega$ being equation of state parameter. Eliminating $\rho$ from the above equations, we get
\beq\label{H1}
2\mathcal{H}'+(3\omega+1)(\mathcal{H}^{2}+1)=\frac{\alpha}{2}(\omega-1)\phi'^{2}+(\omega+1)a^{2} V.
\eeq
From Eq.~(\ref{SF}), we obtain
\beq\label{H2}
\alpha \phi''+2\alpha \mathcal{H} \phi' + a^{2} V_{,\phi}=0.
\eeq

For the Einstein static solutions, the static condition $a'_{0}=a''_{0}=0$ indicates $a=a_{0}=constant$ and $\mathcal{H}_{0}=\mathcal{H}'_{0}=0$. Then, Eq.~(\ref{H1}) reduces to
\beq\label{a0}
a^{2}_{0}=\frac{3\omega_{0}+1}{(\omega_{0}+1)V_{0}}-\frac{\alpha}{2}\frac{(\omega_{0}-1)\phi'^{2}}{(\omega_{0}+1)V_{0}},
\eeq
where the subscript $0$ represents the corresponding value at the Einstein static state. In order to obtain an Einstein static solution, $\omega_{0}$, $\phi'_{0}$ and $V_{0}$ must be constants in the Einstein static state. So, the scalar potential $V$ is flat at the Einstein static state in which $\phi=\phi_{0}$. Considering these conditions, Eq.~(\ref{H2}) becomes
\beq
\frac{dV}{d\phi}|_{\phi=\phi_{0}}=0,
\eeq
which indicates the potential of the scalar field $\phi$ is flat. From Eq.~(\ref{00}), we obtain
\beq
\rho_{0}=\frac{-2(\alpha \phi'^{2}_{0}-2)V_{0}}{2(3\omega_{0}+1)+\alpha(1-\omega_{0})\phi'^{2}_{0}}.
\eeq

Since $a_{0}$ and $\rho_{0}$ are required to be positive, the existence conditions of the Einstein static solutions are $a_{0}^{2}>0$ and $\rho_{0}>0$ which imply
\beq\label{ec1}
\alpha<\frac{2}{\phi'^{2}_{0}}, \quad \omega_{0}>\frac{\alpha \phi'^{2}_{0}+2}{\alpha \phi'^{2}_{0}-6}
\eeq
or
\beq\label{ec2}
\frac{2}{\phi'^{2}_{0}}<\alpha<\frac{6}{\phi'^{2}_{0}}, \quad \omega_{0}<\frac{\alpha \phi'^{2}_{0}+2}{\alpha \phi'^{2}_{0}-6}
\eeq
Here, $\phi'_{0}>0$ and $V_{0}>0$ are taken into consideration.

For a stable Einstein static universe, it requires to be stable against both scalar perturbations and tensor perturbations. In the following, we will discuss the stability of the Einstein static solutions. To simplify this discussion, the tensor perturbations will be studied firstly since they are easy to analyze.

\section{Tensor perturbations}

For the tensor perturbations, the perturbed metric takes the form~\cite{Bardeen1980}
\beq\label{PT}
ds^{2}=-a(\eta)^{2}d\eta^{2}+a(\eta)^{2} (\gamma_{ij}+2h_{ij})dx^{i} dx^{j}.
\eeq
To study the stability of the Einstein static solutions, we perform a harmonic decomposition for the perturbed variable $h_{ij}$
\beq
h_{ij}=H_{T,klm}(t)Y_{ij,klm}(\theta^{n}).
\eeq
Since the quantum numbers $m$ and $l$ do not play a role in the perturbed differential equations, they will be suppressed hereafter. Then, the harmonic function $Y_{k}=Y_{klm}(\theta^{n})$ satisfies~\cite{Harrison1967}
\beq
\Delta Y_{k}=-\mathcal{K}^{2} Y_{k}=-k(k+2)Y_{k}, \quad k=0,1,2,...,
\eeq
where $\Delta$ is the three-dimensional spatial Laplacian operator. Then, substituting the perturbed metric~(\ref{PT}) into the field equations~(\ref{EF}) and using the static conditions, we obtain the equation of tensor perturbations
\beq
H''_{T}+(k^{2}+2)H_{T}=0.
\eeq
For the stable Einstein static solutions, $k^{2}+2>0$ must be satisfied for any $k$. Since $k=0,1,2,...$, it can be seen that the Einstein static solutions are stable against the tensor perturbations in the closed universe.

\section{Scalar perturbations}

In previous section, by considering the tensor perturbation, we find the Einstein static solutions are stable in the closed universe and the tensor perturbations do not constrain any parameter. Since a stable Einstein static solution requires to be stable against both the scalar perturbations and tensor perturbations, we will discuss the stability of the Einstein static solutions against scalar perturbations in the closed universe. Once the Einstein static solutions are stable against both the scalar perturbations and tensor perturbations, the universe can stay at Einstein static state past-eternally. To achieve this goal, we take the perturbed metric in the Newtonian gauge~\cite{Bardeen1980}
\beq\label{PS}
ds^{2}=-a(\eta)^{2}(1+2\Psi)dt^{2}+a(\eta)^{2}(1+2\Phi)\gamma_{ij}dx^{i} dx^{j},
\eeq
where $\Psi$ denotes the Bardeen potential and $\Phi$ represents the perturbation to the spatial curvature. In the Newtonian gauge, the perturbed metric still has the diagonally form and the perturbed variables are gauge invariant. Then, substituting the above perturbed metric~(\ref{PS}) into the field equations~(\ref{EF}) and ~(\ref{SF}), we obtain
\bea
&& 2\nabla^{2}\Phi+6\Phi-\alpha\phi'^{2}_{0}\Psi+\alpha\phi'_{0}\delta\phi'+a^{2}_{0}\delta\rho=0,\label{P1}\\
&& \Phi+\Psi=0,\label{P2}\\
&& 6\Phi''-6\Phi-2\nabla^{2}(\Psi+\Phi)-3\alpha\phi'^{2}_{0}\Psi+3\alpha\phi'_{0}\delta\phi' +3a^{2}_{0}\delta p=0,\label{P3}\\
&& \delta\phi''-\nabla^{2}\delta\phi-\phi'_{0}\dot{\Psi}+3\phi'_{0}\Phi'=0.\label{P4}
\eea
Here, the static conditions and the perturbation of the field $\phi\rightarrow\phi_{0}+\delta\phi$ are used. The relation between density and pressure perturbations is $\delta p=c^{2}_{s}\rho_{0}\delta$ with $\delta=\delta\rho / \rho_{0}$ and $c^{2}_{s}=\omega$.

Similar to discussing the stability of the Einstein static solutions against tensor perturbations, we perform the harmonic decomposition for all perturbed variables
\bea
&& \Psi=\Psi_{k}(\eta)Y_{k}(\theta^{n}), \quad \Phi=\Phi_{k}(\eta)Y_{k}(\theta^{n}),\nonumber\\
&& \delta=\delta_{k}(\eta)Y_{k}(\theta^{n}), \quad \delta\phi=\delta\phi_{k}(\eta)Y_{k}(\theta^{n}).
\eea
Substituting these variables into Eqs.~(\ref{P1}), ~(\ref{P2}), ~(\ref{P3}), ~(\ref{P4}) and eliminating $\Phi_{k}$ and $\delta_{k}$, we get two independent perturbed equations
\bea
&& 2\Phi''_{k}+\Big[2(\omega_{0} k^{2}-3\omega_{0}-1)+\alpha(1-\omega_{0}) \phi'^{2}_{0} \Big]\Phi_{k}+\alpha(1-\omega_{0})\phi'_{0}\delta\phi'_{k}=0,\label{TP1}\\
&& \delta\phi''_{k}+k^{2}\delta\phi_{k}+4\phi'_{0} \Phi'_{k}=0.\label{TP2}
\eea

Introducing two new variables $A_{k}=\Phi'_{k}$ and $B_{k}=\delta\phi'_{k}$, the perturbed equations ~(\ref{TP1}) and ~(\ref{TP2}) can be reduced as
\bea
&& A'_{k}+a_{11}B_{k}+a_{12}\Phi_{k}=0,\\
&& B'_{k}+b_{11}A_{k}+b_{12}\delta\phi_{k}=0,\\
&& \Phi'_{k}-A_{k}=0,\\
&& \delta\phi'_{k}-B_{k}=0,
\eea
with
\bea
&& a_{11}=\frac{1}{2}\alpha(1-\omega_{0})\phi'_{0},\\
&& a_{12}=\frac{1}{2}\alpha(1-\omega_{0})\phi'^{2}_{0}+\omega_{0}k^{2}-3\omega_{0}-1,\\
&& b_{11}=4\phi'_{0},\\
&& b_{12}=k^{2},
\eea

The stability of the Einstein static solutions can be determined by the eigenvalues of the coefficient matrix of this dynamical system, which are
\bea
\mu^{2}=\frac{-M\pm\sqrt{N}}{2},
\eea
where
\bea
&& M=a_{12}-a_{11}b_{11}+b_{12},\\
&& N=(a_{12}-a_{11}b_{11}+b_{12})^{2}-4a_{12}b_{12}.
\eea
For $\mu^{2}>0$, one perturbation from the Einstein static state will leads to an exponential deviation from the Einstein static state, and the corresponding Einstein static solution is unstable. While for $\mu^{2}<0$, a small perturbation from the Einstein static state will result in an oscillation around this state, and the corresponding Einstein static state is stable, which means that it is stable both in the past and in the future. Thus, under the scalar perturbations, the stability conditions are given by $\mu^{2}<0$, which indicate
\bea\label{SCS}
M>0, \quad N>0, \quad M^{2}-N>0.
\eea

\begin{figure*}[htp]
\begin{center}
\includegraphics[width=0.45\textwidth]{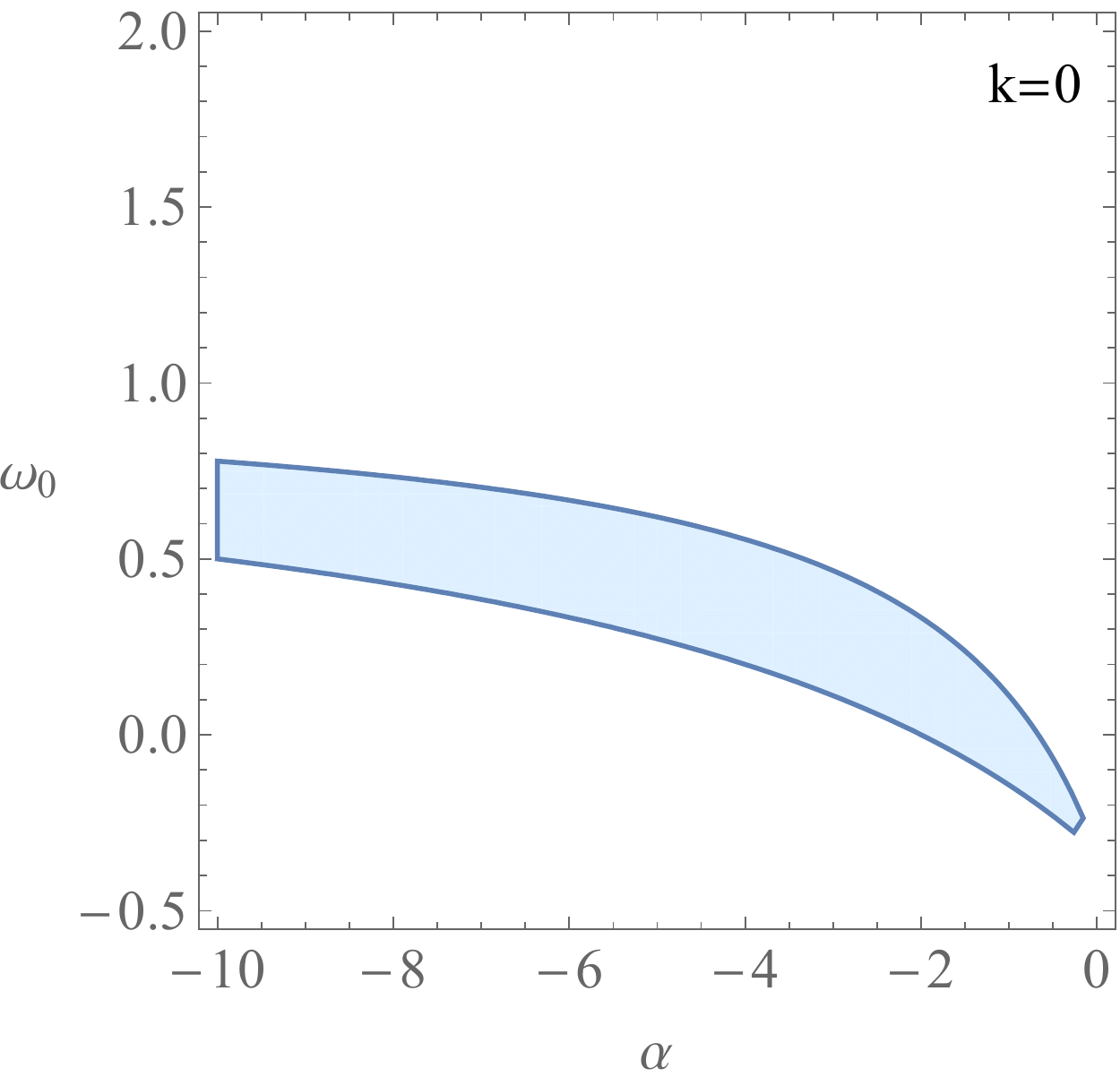}
\includegraphics[width=0.45\textwidth]{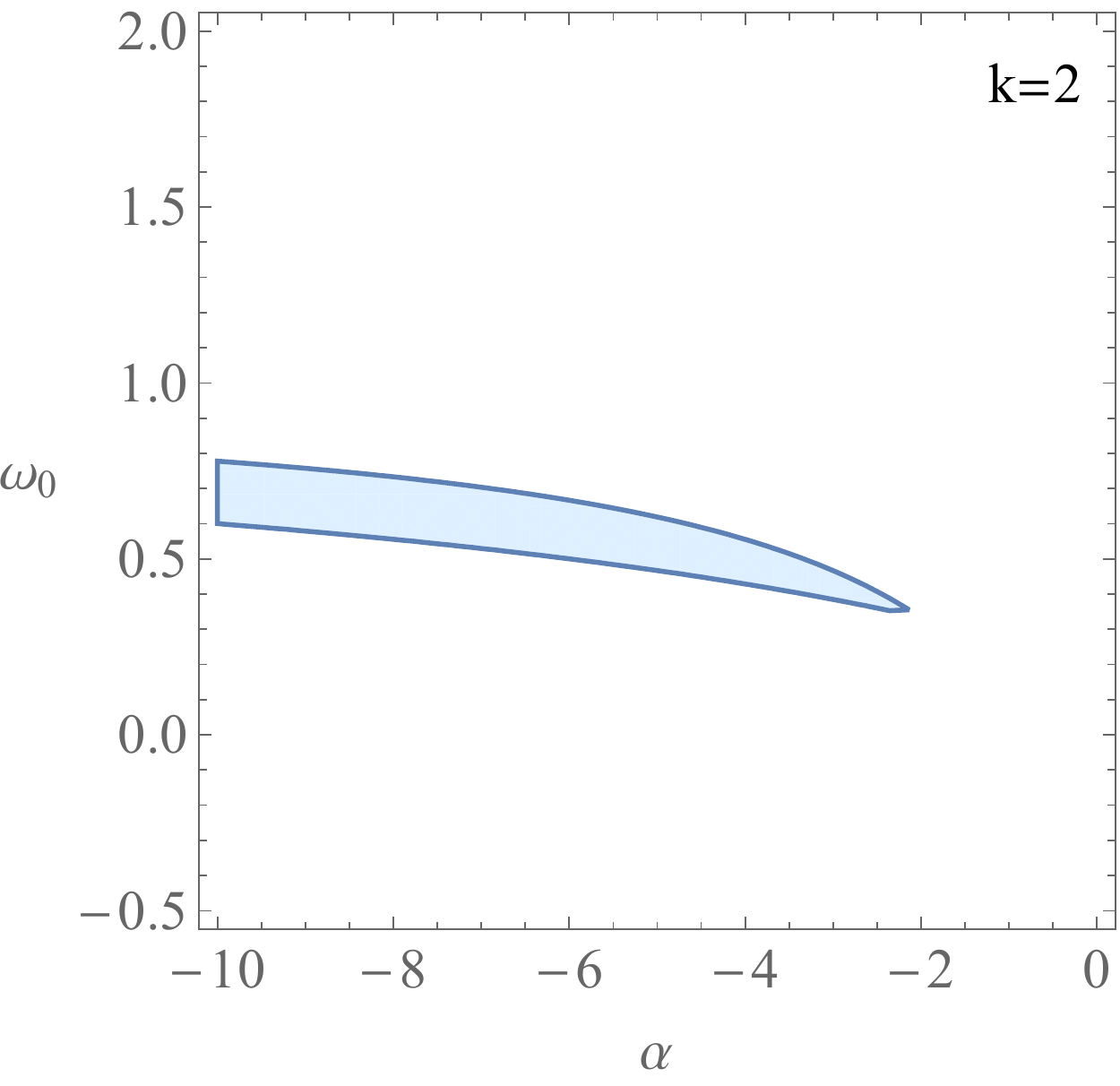}
\includegraphics[width=0.45\textwidth]{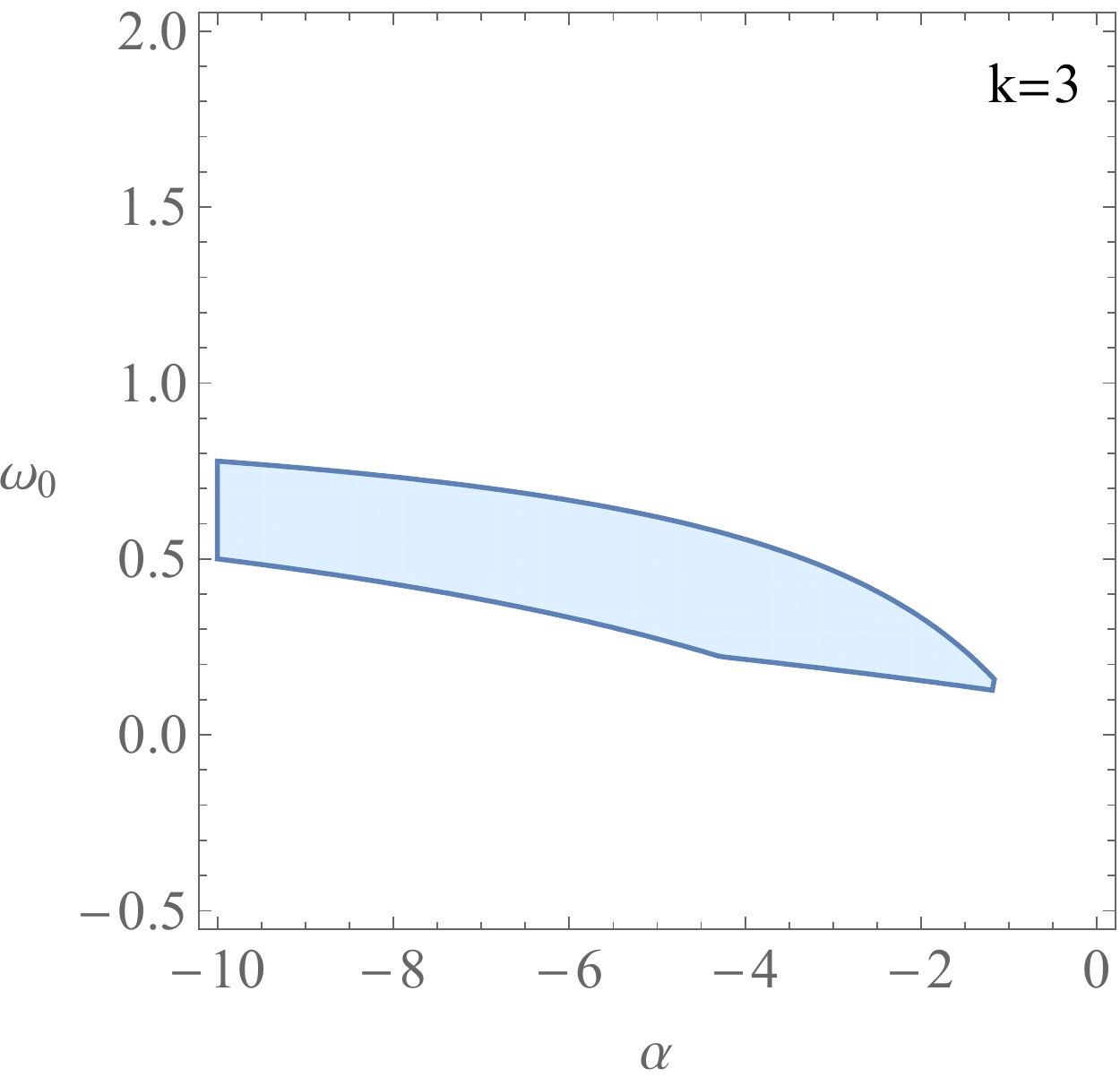}
\includegraphics[width=0.45\textwidth]{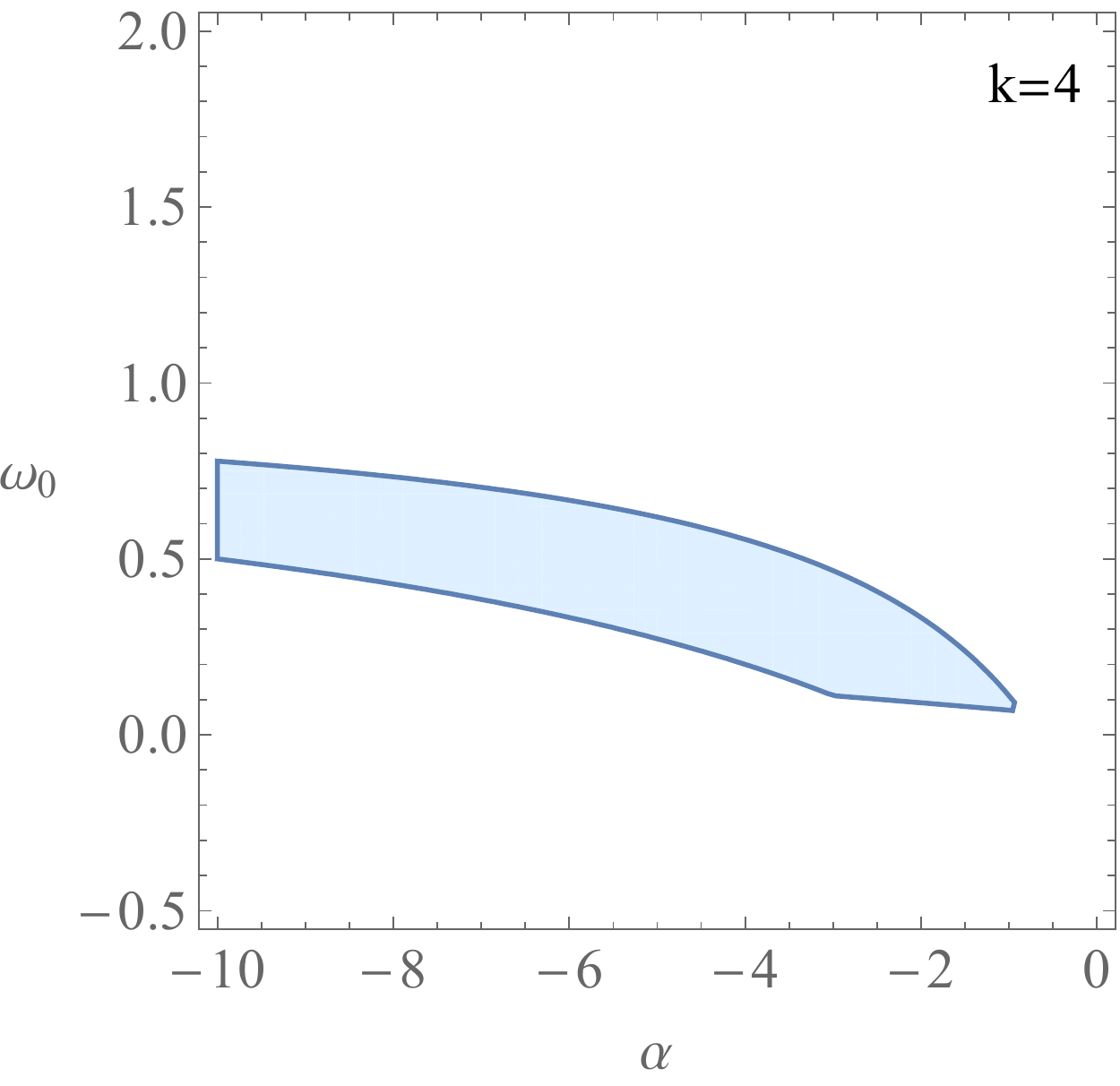}
\caption{\label{Fig1} Stability regions in $(\omega_{0},\alpha)$ plane under homogeneous and inhomogeneous scalar perturbations with $k=0,2,3,4$. $k=0$ represents the homogeneous scalar perturbations. These figures are plotted for $\phi'_{0}=1$ and $V_{0}=5$.}
\end{center}
\end{figure*}

Since the homogeneous scalar perturbations correspond to the case $\mathcal{K}^{2}=0$, we obtain $b_{12}=0$ and $M^{2}=N$. Therefore, the stability conditions for the homogeneous scalar perturbations reduce as $M>0$ which require
\bea
\phi'>0, V>0, \alpha<0, \frac{\alpha\phi'^{2}_{0}+2}{\alpha\phi'^{2}_{0}-6}<\omega_{0}<\frac{3\alpha\phi'^{2}_{0}+2}{3\alpha\phi'^{2}_{0}-6}.
\eea
Here, the existence conditions Eqs.~(\ref{ec1}) and ~(\ref{ec2}) are taken into consideration. The stability region for homogeneous scalar perturbations is shown in the first panel of Fig.~(\ref{Fig1}).

For the inhomogeneous scalar perturbations, Since $k=1$ mode represents a gauge degree of freedom corresponding to a global rotation, the modes have $k \geq 2$ which indicates $\mathcal{K}^{2} \geq 8$.

Since the Einstein static solutions must be stable against all kinds of perturbations, we will discuss the stability conditions against inhomogeneous scalar perturbations under the existence conditions and the stability conditions for the homogeneous scalar perturbations. Thus, the stability conditions are given by equation~(\ref{SCS}) which indicate
\bea\label{stable}
\phi'>0, V>0, \alpha<-\frac{2}{\phi'^{2}_{0}}, \frac{\alpha\phi'^{2}_{0}-2}{\alpha\phi'^{2}_{0}-10}<\omega_{0}<\frac{3\alpha\phi'^{2}_{0}+2}{3\alpha\phi'^{2}_{0}-6}.
\eea

In Fig.~(\ref{Fig1}), we plot some examples of the stable regions of the homogeneous and inhomogeneous scalar perturbations. In this figure, the values of $k$ are taken to be $k=0,2,3,4$, and $k=0$ represents the homogeneous scalar perturbations. For the inhomogeneous scalar perturbations, the stable regions become larger and larger with the increase of the value of $k$. It is obvious that the smallest stability region is obtained in the case $k=2$.

\begin{figure*}[htp]
\begin{center}
\includegraphics[width=0.45\textwidth]{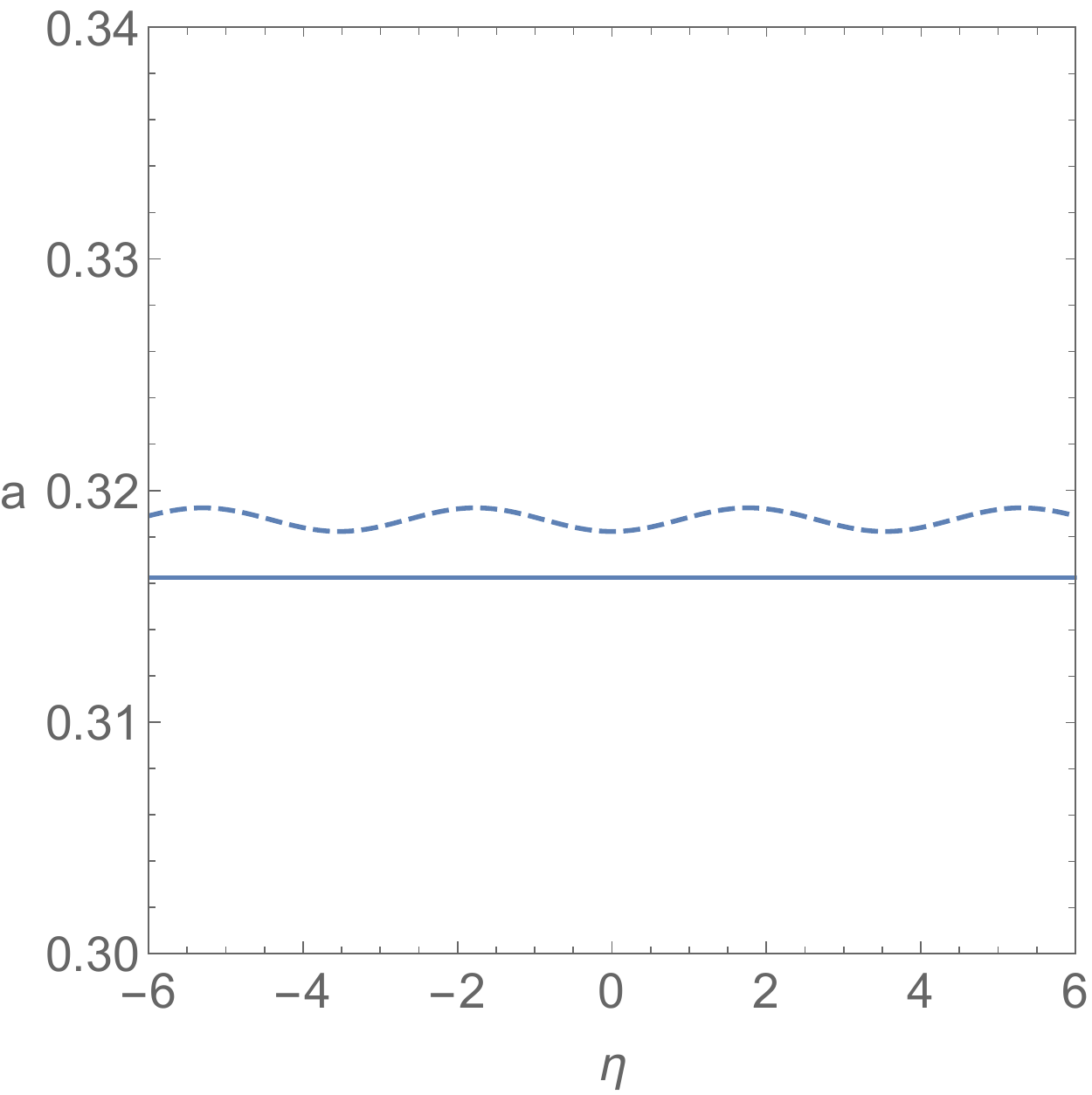}
\includegraphics[width=0.45\textwidth]{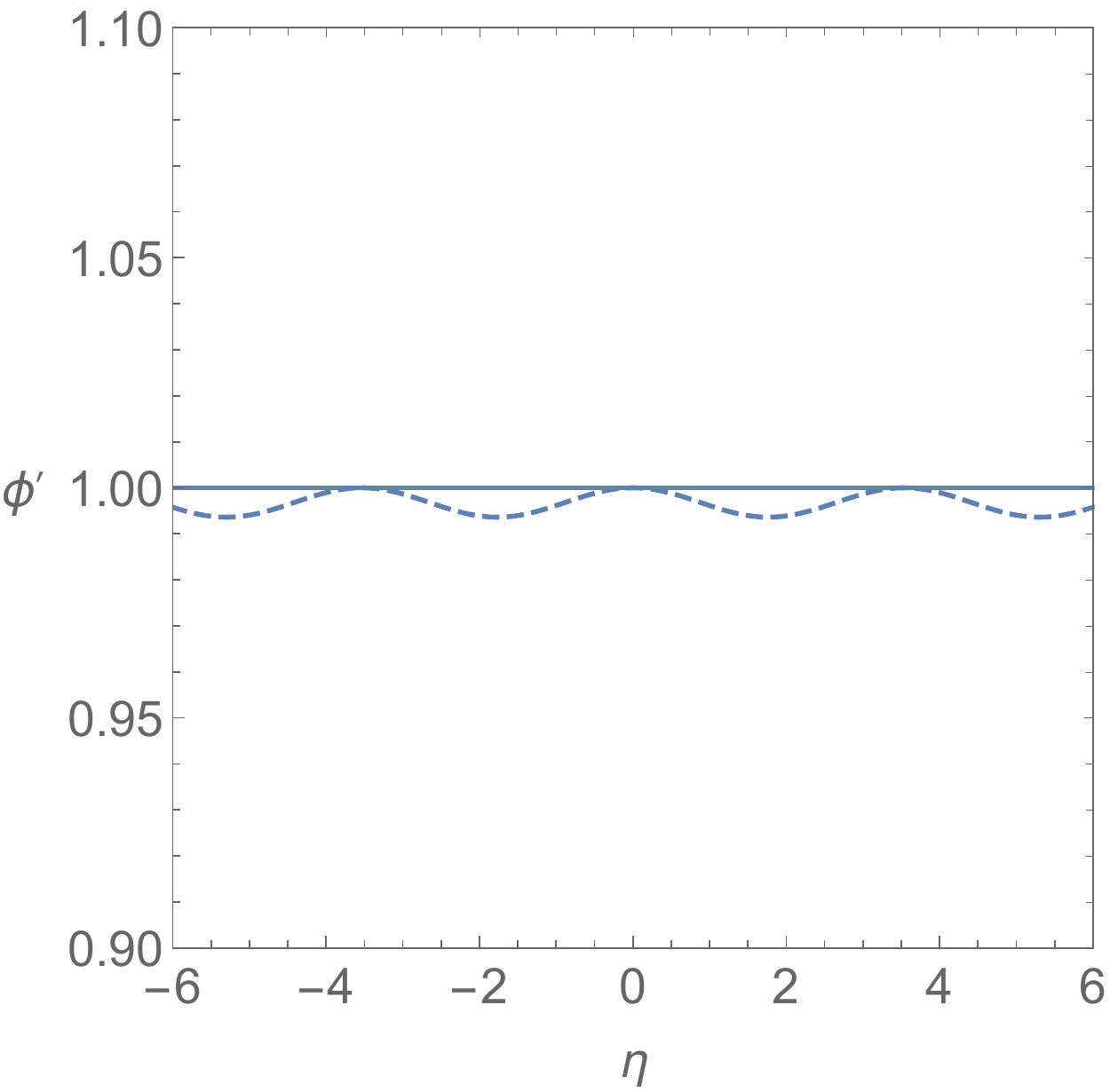}
\caption{\label{Fig2} Evolutionary curves of $a$ and $\phi'$ with $\eta$. The solid line represents that the initial value of $a$ is the Einstein static solution $a_{0}$, while the dashed line denotes that the initial value of $a$ has a slight deviation of $a_{0}$. These figures are plotted for $\phi'_{0}=1$, $V_{0}=5$, $\alpha=-10$ and $\omega_{0}=0.6$.}
\end{center}
\end{figure*}

In Fig.~(\ref{Fig2}), the evolutionary curves of $a$ and $\dot{\phi}$ against $\eta$ are depicted. In this figure, the solid line represents that the initial value of $a$ is the Einstein static solution $a_{0}$, while the dashed line denotes that the initial value of $a$ has a slight deviation of $a_{0}$. When the initial value of $a$ is chosen as the Einstein static solution $a_{0}$, it can be seen that the evolutionary curves of $a$ and $\dot{\phi}$ parallel to $\eta$-axis. Considering a slight deviation of the initial value $a_{0}$, the evolutionary curves of $a$ and $\dot{\phi}$ oscillate near the Einstein static state, which are depicted by the dashed line in Fig.~(\ref{Fig2}). These figures show that the Einstein static solutions are stable under the stability conditions ~(\ref{stable}).

In Ref.~\cite{Barrow2003}, the Einstein static universe was studied in general relativity with quintessence, and it is unstable against inhomogeneous perturbations. Comparing with quintessence, the action~(\ref{action}) in this paper has a coupling parameter $\alpha$ and the case $\alpha=1$ corresponds to quintessence. Due to the presence of parameter $\alpha$, the stability conditions are extended, so that a wider range of Einstein static solutions can be obtained. It is found that the Einstein static universe is stable against both scalar and tensor perturbations under conditions ~(\ref{stable}) which suggest that a stable Einstein static solution requires a negative $\alpha$.

\section{Leaving the Einstein static state}

In previous section, we find that the Einstein static solutions are stable, which indicates the universe can stay at Einstein static state past-eternally. In the emergent universe, we require the universe can exit from the stable Einstein static state and enter into an inflationary era. In this section, we will discuss how to realize this transition.

\begin{figure*}[htp]
\begin{center}
\includegraphics[width=0.45\textwidth]{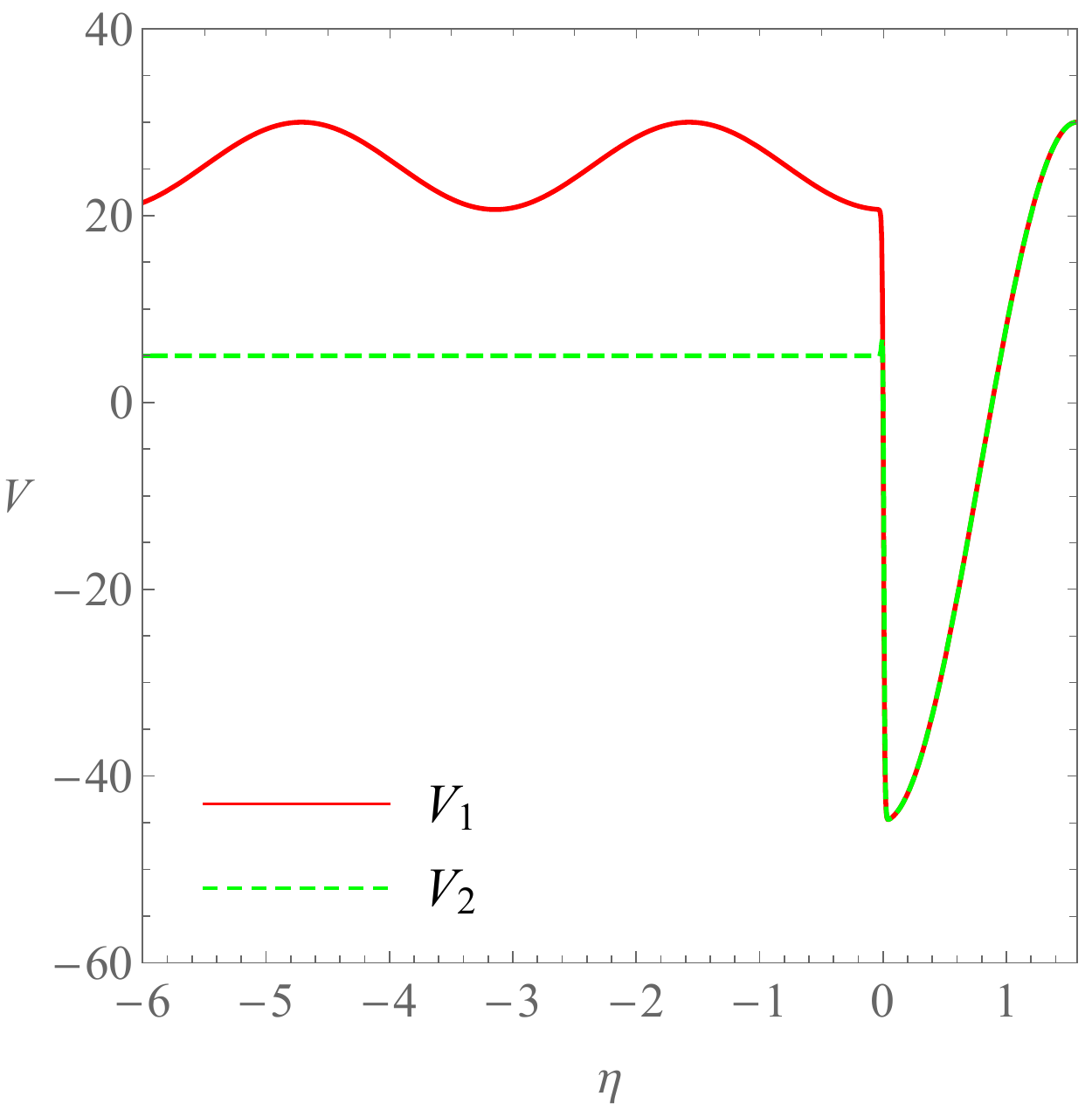}
\includegraphics[width=0.46\textwidth]{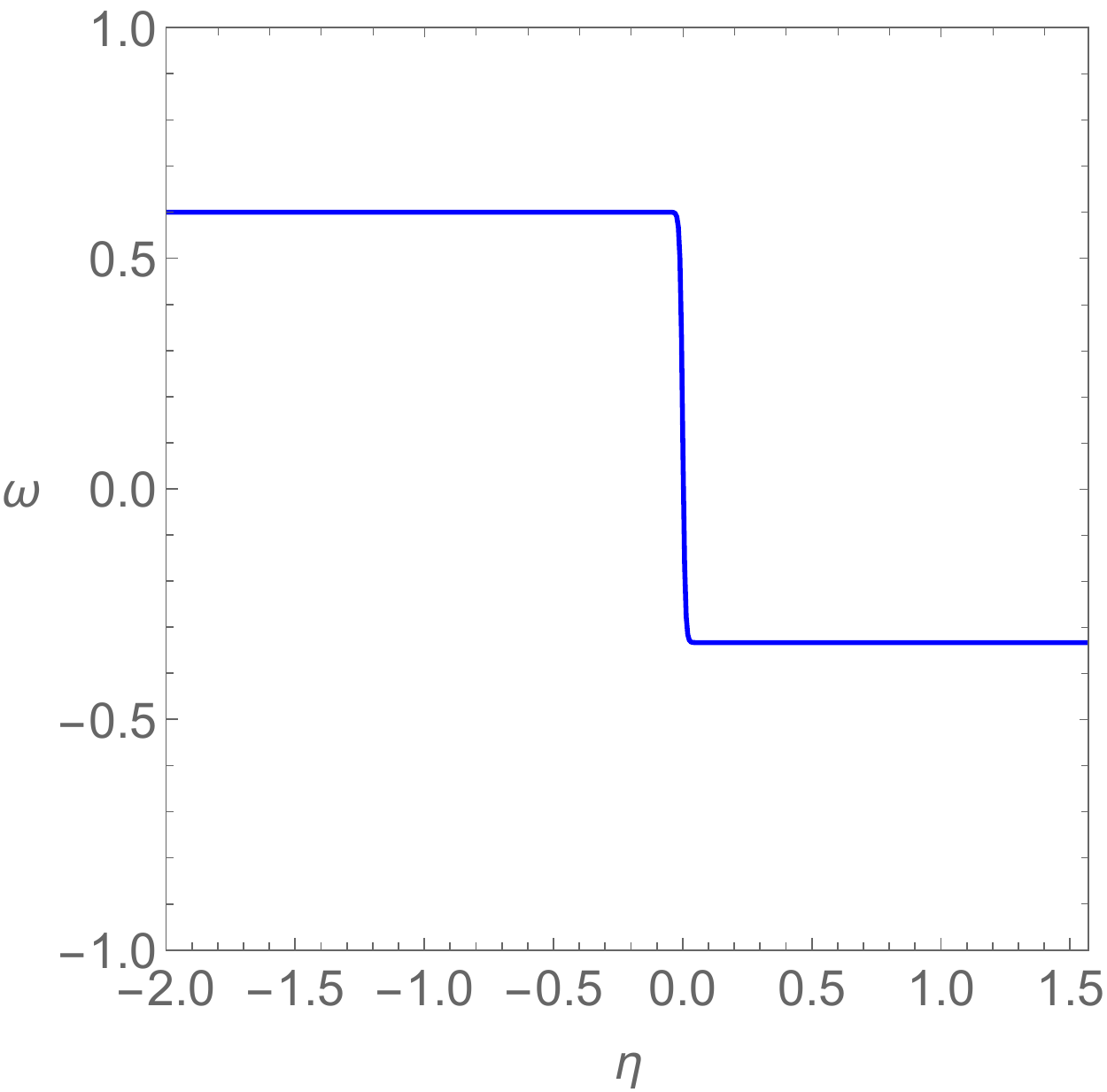}
\includegraphics[width=0.46\textwidth]{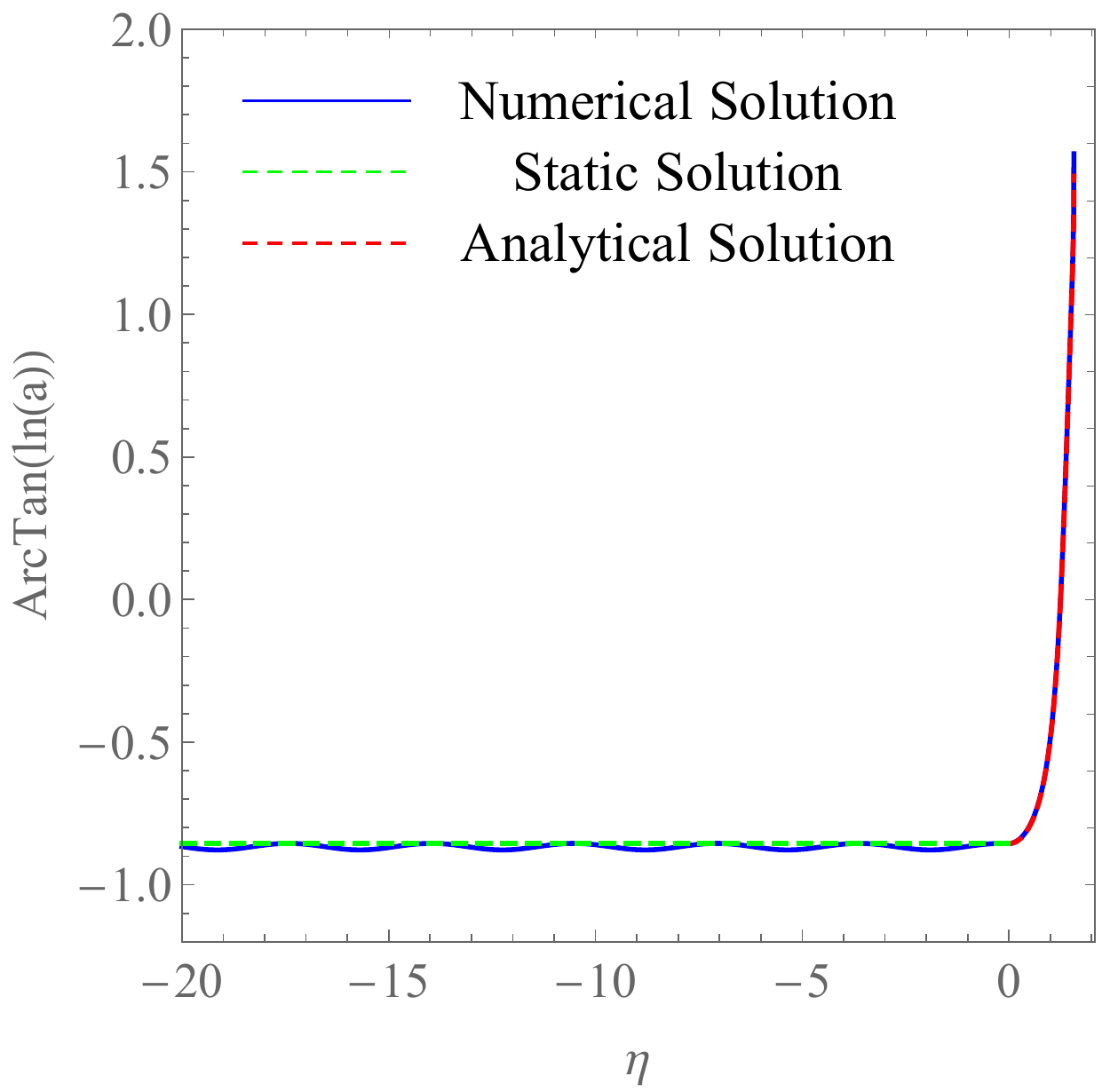}
\includegraphics[width=0.45\textwidth]{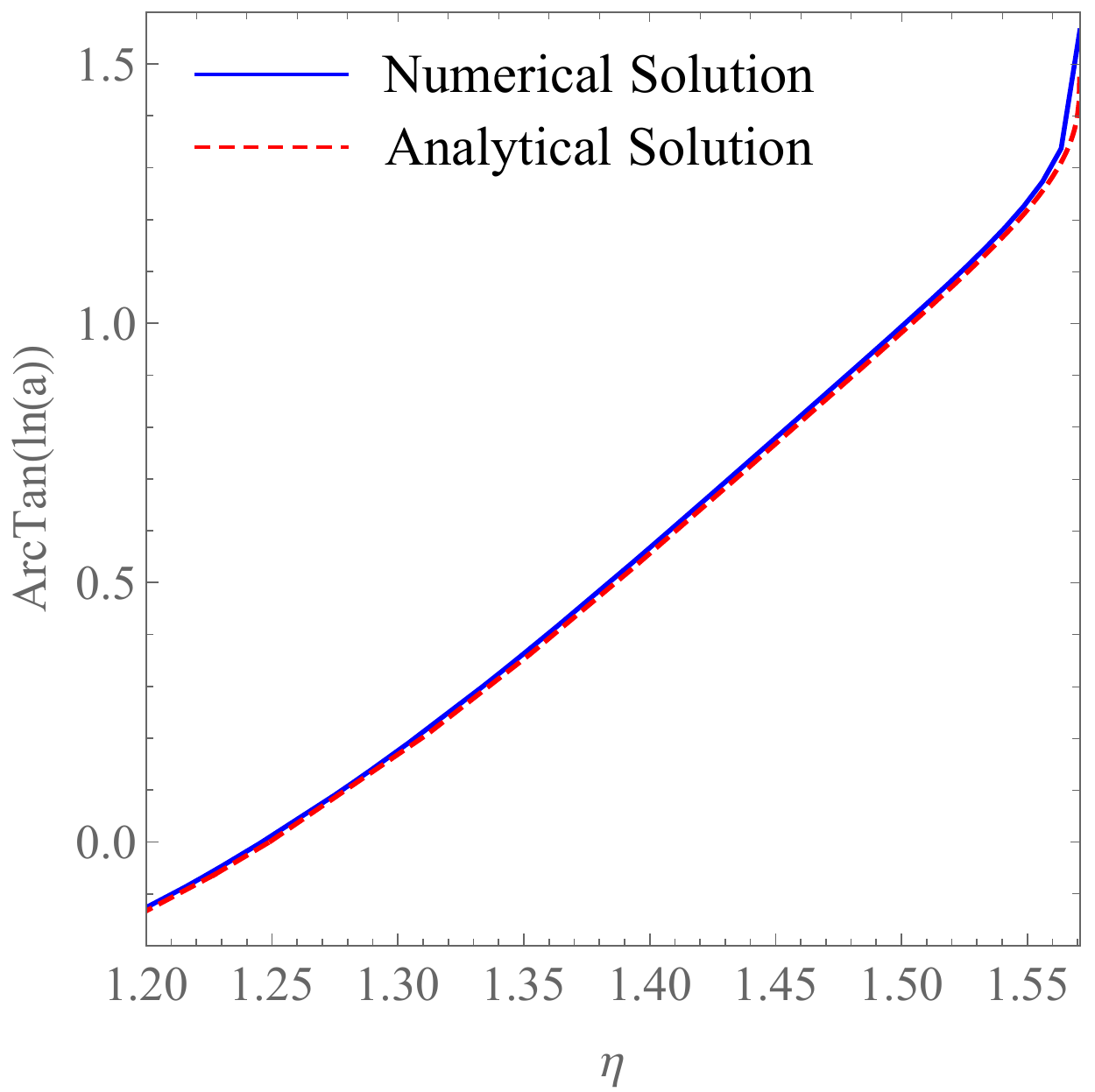}
\caption{\label{Fig3} Evolutionary curves of $V$, $\omega$ and $a$ against $\eta$ with $\phi'=1$, $V_{0}=5$, $\alpha=-10$, $\omega=0.6$ and $\xi=100$.}
\end{center}
\end{figure*}

In the Einstein static state, since $a_{0}$, $\phi'_{0}$ and $V_{0}$ are constant, $a_{0}$ is given in Eq.~(\ref{a0}) and the scalar potential $V_{0}$ is flat, we can take $\phi=\beta(\eta-\eta_{t})$. Here, $\beta$ is a constant and $\eta_{t}$ denotes the transition time that the universe exits from the Einstein static state. After the universe exits from the Einstein static state, it evolves into the inflationary era. In the inflationary era, the universe is dominated by the scalar field $\phi$, and the effect of the perfect fluid is negligible. By ignoring $\rho$ and $p$ in Eqs.~(\ref{00}) and ~(\ref{ij}) and considering $\phi'^{2} \ll a^{2}V(\phi)$ in the slow-roll inflation stage, we eliminate $\phi'^{2}$ in Eqs.~(\ref{00}) and ~(\ref{ij}). Then, combining Eqs.~(\ref{00}) and ~(\ref{ij}) to eliminating the scalar potential $V$, we get
\beq
\mathcal{H}'-\mathcal{H}^{2}-1=0,
\eeq
with the solution
\beq\label{ai}
a=\frac{a_{0}}{\cos(\eta-\eta_{t})}
\eeq
which is obtained in Ref.~\cite{Shumaylov2022}, and it can also be obtained by solving Eqs.~(\ref{H1}) or ~(\ref{H2}). Assuming the scalar potential takes the form
\beq\label{V1}
V_{1}=\frac{3}{a^{2}_{0}}+\frac{\alpha \beta^{2} (1-\omega)\cos^{2}(\phi/\beta)}{2(1+\omega)a^{2}_{0}}
\eeq
with $\omega=-1/3$, and then solving Eq.~(\ref{H1}) or Eq.~(\ref{H2}), we find that both Eqs.~(\ref{H1}) and ~(\ref{H2}) can give the solution ~(\ref{ai}). An example of $V_{1}$ is plotted by the red line in the first panel of Fig.~(\ref{Fig3}).  As a result, the expression of scale factor $a$ can be written as follows
\bea
a(\eta)=\Bigg\{
\begin{array}{ll}
a_0,\quad & \eta<\eta_t\\
\frac{a_0}{\cos(\eta-\eta_t)},\quad & \eta_t \leq \eta < \eta_t+\frac{\pi}{2}.\label{Comp}
\end{array}
\eea

In order to realize the exit of the universe from the stable Einstein static state and evolve into the inflationary era, it is necessary to break the stability conditions of the Einstein static state. To achieve this goal, we need to construct a scalar potential $V$ and a equation of state parameter $\omega$ to realize this transition. According to previous discussion, both $V$ and $\omega$ are constants in the Einstein static state, while $V$ has the form in Eq.~(\ref{V1}) and $\omega$ takes the value $-1/3$ in the slow-roll inflationary era. So, we require that the scalar potential $V$ and the equation of state parameter $\omega$ vary with the conformal time $\eta$ slowly, they approach to a constant in the Einstein static state $\eta \rightarrow -\infty$ and decrease rapidly when inflation begins $\eta \rightarrow \eta_{t}$. Considering these conditions, we construct a scalar potential
\bea\label{V2}
V_{2}=&&\frac{1}{2}\Big[1-\tanh\big(\frac{\xi}{\beta}\phi\big)\Big]V_{0}\nonumber\\
&&+\frac{1}{2}\Big[1+\tanh\big(\frac{\xi}{\beta}\phi\big)\Big]\Big[\frac{3}{a^{2}_{0}}+\frac{\alpha \beta^{2} (1-\omega)\cos^{2}(\frac{\phi}{\beta})}{2(1+\omega)a^{2}_{0}}\Big]\nonumber\\
\eea
and a equation of state parameter
\beq\label{w1}
\omega=\frac{1}{2}\big[1-\tanh(\xi(\eta-\eta_{t}))\big]\omega_{0}+\frac{1}{2}\big[1+\tanh(\xi(\eta-\eta_{t}))\big]\omega_{1}
\eeq
with $\omega_{1}=-1/3$.

An example of $V_{2}$ and $\omega$ is shown in the upper panels of Fig.~(\ref{Fig3}). The potential $V_{2}$ (Eq.~(\ref{V2})) is plotted by the green dashed line in the first panel, while the equation of state parameter $\omega$ (Eq.~(\ref{w1})) is plotted in the second panel. With the time passes and approaches to the transition time $0$, the potential $V_{2}$ deviates from the static value $V_{0}$ and decreases to less than $0$, and $\omega$ deviates from the static value $\omega_{0}$ and decreases to $-1/3$. Thus, the stability conditions for the Einstein static state are broken, the universe exits from the stable Einstein static state and inflation begins.

Using the expression of $V_{2}$ (Eq.~(\ref{V2})) and $\omega$ (Eq.~(\ref{w1})), we solve the dynamical equations ~(\ref{H1}) and ~(\ref{H2}) numerically and depict the evolution of universe in the early time in the lower panels of Fig.~(\ref{Fig3}). In these two panels, the static value of scale factor $a_{0}$ (Eq.~(\ref{a0})) is plotted by green dashed line, the scale factor $a$ in the inflationary era (Eq.~(\ref{ai})) is shown by red dashed line, and the scale factor depicted by blue line is the numerical simulation result obtained by solving the dynamical equations ~(\ref{H1}) and ~(\ref{H2}) numerically. For convenience, we adopt $\eta=0$ as the time at which the universe exits from the Einstein static state and the transition time takes the value $\eta_{t}=0$. In the case of Fig.~(\ref{Fig3}), when the scalar potential decreases to a negative value, the stability condition is broken and the universe exits from the Einstein static state, and then the potential bounces back and becomes positive again. Then, inflation begins at the time when the potential becomes positive. It can be seen from these figures in the lower panels that the numerical and analytical solutions of the Friedmann equations overlap and the evolution of scale factor $a$ can be described by Eq.~(\ref{Comp}). And the right panel depicts that, after the universe leaves the initial Einstein static state, the factor $a$ increases rapidly and the universe evolves into the inflationary era. Thus, the emergent universe can be realized successfully and the big bang singularity can be avoided in general relativity with k-essence.

\section{CMB power spectrum}

In previous section, we find that the Einstein static universe can be stable against tensor and scalar perturbations in k-essence, and can exit the stable static state and then evolve into a subsequently inflation era. In this section, we will analyze the CMB TT-spectrum for the emergent universe and discuss whether the suppression of CMB TT-spectrum can be realized in this successful emergent universe.

To study the primordial power spectrum in emergent universe, we introduce a gauge invariant comoving curvature perturbation
\beq
\mathcal{R}=\Psi+\frac{\mathcal{H}}{\phi'}\delta\phi
\eeq
which satisfies the curved Mukhanov-Sasaki equation in k-essence~\cite{Shumaylov2022}
\beq\label{MS}
v''_{k}+\Big(c^{2}_{s}\mathcal{K}^{2}-\frac{\mathcal{Z}''}{\mathcal{Z}}-\frac{2}{\mathcal{H}}\frac{\mathcal{Z}'}{\mathcal{Z}}+1-3c^{2}_{s}\Big)v_{k}=0
\eeq
with
\bea
&& v=\mathcal{Z}\mathcal{R}, \quad \mathcal{Z}=z\sqrt{\frac{\mathcal{D}^{2}}{\mathcal{D}^{2}-\varepsilon}},\nonumber\\
&& z=\frac{a^{2}(\rho_{\phi}+p_{\phi})^{1/2}}{\mathcal{H}c_{s}}, \quad \varepsilon=\frac{a^{2}(\rho_{\phi}+p_{\phi})}{2\mathcal{H}^{2}c^{2}_{s}},\nonumber\\
&& \mathcal{D}^{2}=-k(k+2)+3.
\eea
Here, $\rho_{\phi}$ and $p_{\phi}$ denote the energy density and pressure for scalar field, and the effect of perfect fluid is not taken into account since $\rho_{\phi}$ dominates the evolution of the universe before inflation ends. $c^{2}_{s}$ denotes the squared sound speed of inflation which is given as
\beq
c^{2}_{s}=\frac{P_{,X}}{P_{,X}+2X P_{,XX}}.
\eeq
For the case $P(X,\phi)=\alpha X-V(\phi)$, we obtain $c^{2}_{s}=1$. Then, the curved Mukhanov-Sasaki equation ~(\ref{MS}) reduces to that in Ref.~\cite{Thavanesan2021, Huang2022}.

In the Einstein static state, considering the static condition, we get
\beq
\mathcal{Z}=a_{0}\sqrt{2[k(k+2)-3]} \sim a_{0},
\eeq
and Eq.~(\ref{MS}) becomes
\beq\label{MS1}
v''_{k}+k^{2}_{-} v_{k}=0, \quad k^{2}_{-}=k(k+2)-4,
\eeq
which has the solution
\beq
v_{k}(\eta)=A_{k}e^{ik_{-}\eta}+B_{k}e^{-ik_{-}\eta}.
\eeq
Using the normalization conditions and choosing the Bunch-Davies vacuum, we get the initial condition
\beq
A_{k}=0, \quad B_{k}=\sqrt{\frac{1}{2k_{-}}}.
\eeq
So, the solution of Eq.~(\ref{MS1}) is determined as
\beq\label{vk0}
v_{k}(\eta)=\sqrt{\frac{1}{2k_{-}}}e^{-ik_{-}\eta}.
\eeq

In the inflationary era, since $\varepsilon \approx 0$, we get
\beq\label{55}
\frac{\mathcal{Z}''}{\mathcal{Z}}+\frac{2}{\mathcal{H}}\frac{\mathcal{Z}'}{\mathcal{Z}}+2 \approx \frac{a''}{a}+3.
\eeq
Then, substituting Eqs.~(\ref{ai}) and ~(\ref{55}) into Eq.~(\ref{MS}), the curved Mukhanov-Sasaki equation ~(\ref{MS}) becomes
\beq\label{MS2}
v''_{k}+\Big[k^{2}_{+}-\frac{2}{(\eta-(\eta_{t}+\frac{\pi}{2}))^{2}}\Big]v_{k}=0, \quad k^{2}_{+}=k(k+2)-\frac{8}{3}.
\eeq
The solution of the above equation takes the form
\bea\label{vkcd}
v_k(\eta)=\sqrt{\frac{\pi}{4}}\sqrt{\big(\eta_t+\frac{\pi}{2}\big)-\eta}\Big[C_k H^{(1)}_{3/2}\Big(k_{+}\big((\eta_t+\frac{\pi}{2})-\eta\big)\Big)+D_k H^{(2)}_{3/2}\Big(k_{+}\big((\eta_t+\frac{\pi}{2})-\eta\big)\Big)\Big],\nonumber\\
\eea
where $H^{(1)}$ and $H^{(2)}$ are the Hankel functions of the first and second kinds, and $C_{k}$ and $D_{k}$ are integration constants. In order to determine $C_{k}$ and $D_{k}$, we match Eqs.~(\ref{vk0}) and ~(\ref{vkcd}) at the transition time $\eta_{t}$ by using the continuity condition of $v_{k}$ and $v'_{k}$, then we obtain
\bea
&&C_k=\frac{1}{4}e^{-i k_{-} \eta_t}\sqrt{\frac{1}{k_{-}}}\Big[i \pi k_{+} H^{(2)}_{1/2}\Big(\frac{\pi}{2}k_{+}\Big)+(-2i+\pi k_{-})H^{(2)}_{3/2}\Big(\frac{\pi}{2}k_{+}\Big)\Big],\\
&&D_k=-\frac{1}{4}e^{-i k_{-} \eta_t}\sqrt{\frac{1}{k_{-}}}\Big[i \pi k_{+} H^{(1)}_{1/2}\Big(\frac{\pi}{2}k_{+}\Big)+(-2i+\pi k_{-})H^{(1)}_{3/2}\Big(\frac{\pi}{2}k_{+}\Big)\Big].
\eea

The curved primordial power spectrum of the comoving curvature perturbation $\mathcal{R}$ is defined as
\beq\label{PR}
\mathcal{P}_{\mathcal{R}}=\frac{k^3}{2\pi^2}\left| \mathcal{R}_k \right|^2 =\frac{k^3}{2\pi^2}\left| \frac{v_k}{\mathcal{Z}_k} \right|^2.
\eeq
Substituting Eq.~(\ref{vkcd}) into Eq.~(\ref{PR}), we get the curved primordial power spectrum of $\mathcal{R}$
\bea
\mathcal{P}_{\mathcal{R}} && =\frac{k^3}{2\pi^2}\left| \mathcal{R}_k \right|^2 \nonumber\\
&& \approx \lim_{\eta\rightarrow\eta_t+\frac{\pi}{2}}\frac{1}{8a^{2}\pi^{2}\varepsilon}\frac{1}{\big[\eta-\big(\eta_t+\frac{\pi}{2}\big)\big]^2}\frac{k^3}{k^3_{+}} \left| C_k-D_k \right|^2\nonumber\\
&& =A_s \frac{k^3}{k^3_{+}}\left| C_k-D_k \right|^2.
\eea
in which the transition time parameter $\eta_t$, slow-roll parameter $\varepsilon$, and formally diverging parameters are absorbed into the scalar power spectrum amplitude $A_s$ ~\cite{Thavanesan2021}.

Then, the analytical primordial power spectrum can be parameterized as
\bea
\mathcal{P}_{\mathcal{R}}(k)=A_{s}\Big(\frac{k}{k_{*}}\Big)^{n_s-1}\frac{k^3}{k^3_{+}}\left| C_k-D_k \right|^2,
\eea
where $k_{*}=0.05 Mpc^{-1}$ represents the pivot perturbation mode.

\begin{figure*}[htp]
\begin{center}
\includegraphics[width=0.45\textwidth]{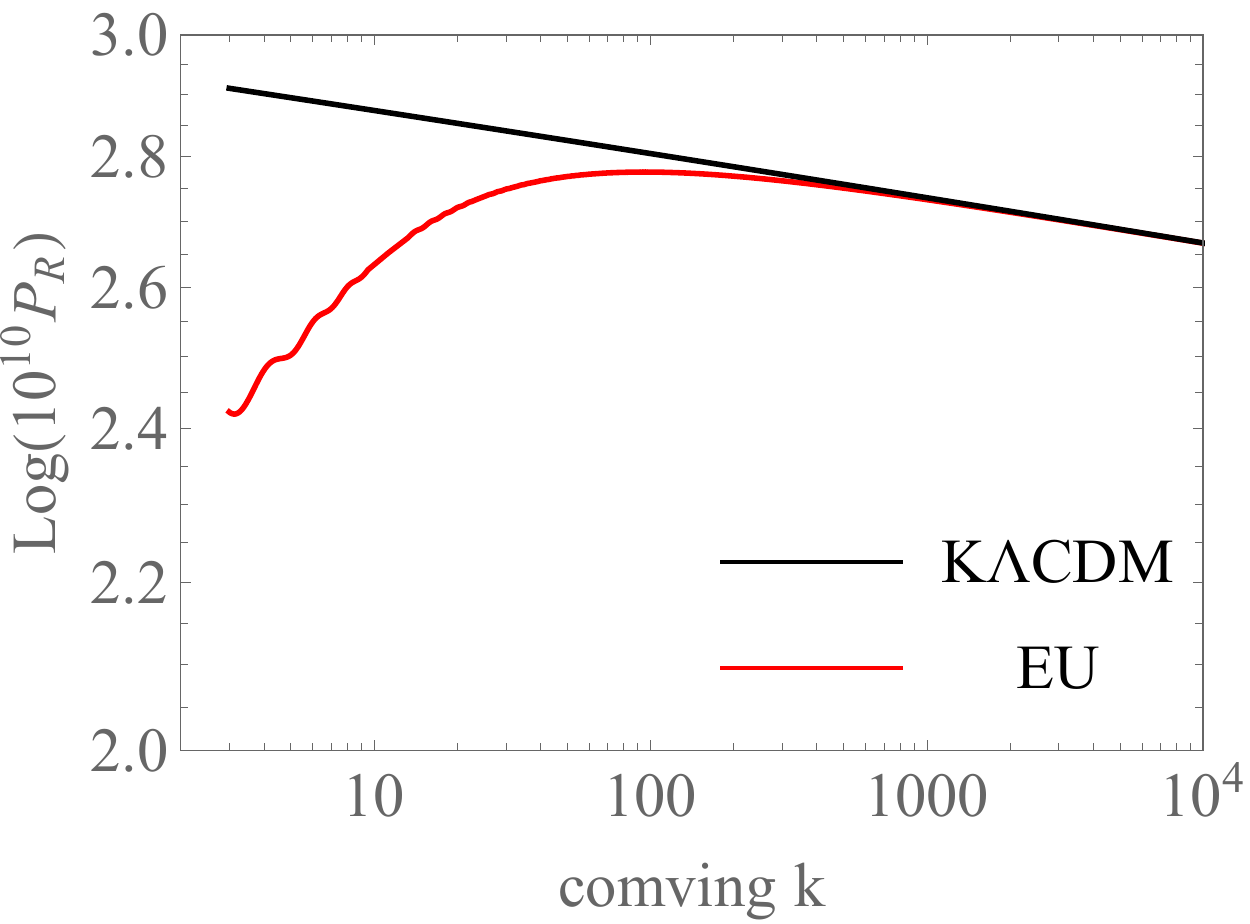}
\includegraphics[width=0.45\textwidth]{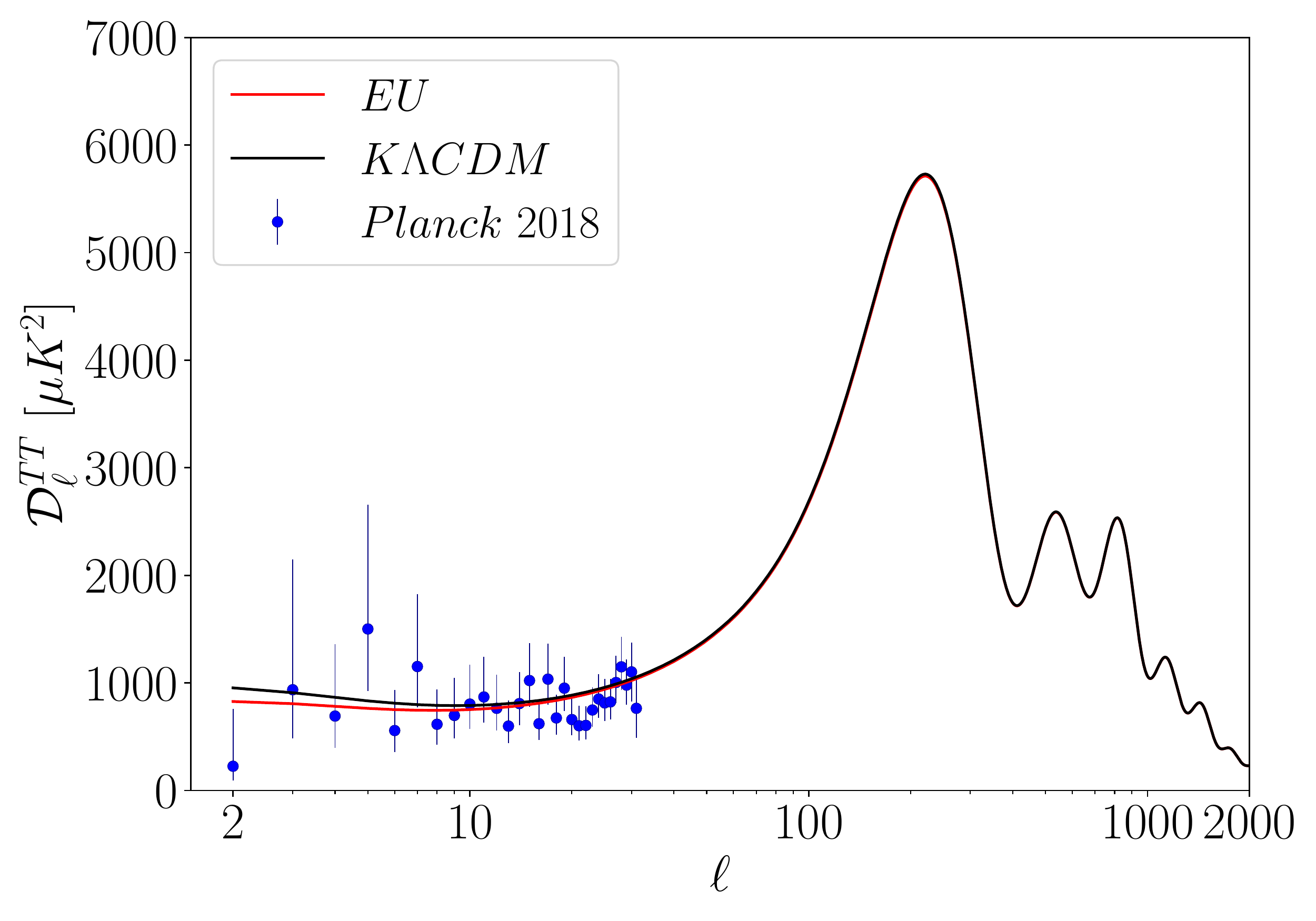}
\caption{\label{Fig4} Primordial power spectrum and CMB TT-spectrum for the emergent universe(EU).}
\end{center}
\end{figure*}

To depict the primordial power spectrum, we use the Planck 2018 results in the curved universes best-fit data (TT,TE,EE+lowl+lowE+lensing) $A_s=2.0771 \pm 0.1017 \times 10^{-9}$ and $n_s = 0.9699 \pm 0.0090$. In the left panel of Fig.~(\ref{Fig4}), we have plotted the primordial power spectrum for the emergent universe. The red line denotes the primordial power spectrum of the emergent universe, while the black one corresponds to the one of $\Lambda$CDM with positive spatial curvature and we label it as K$\Lambda$CDM. The left panel of Fig.~(\ref{Fig4}) shows that the spectrum is suppressed for $k<30$. Then, using CLASS code~\cite{Blas2011}, we have plotted the CMB TT-spectrum which is shown in the right panel of Fig.~(\ref{Fig4}). From this figure, we can see that the CMB TT-spectrum of the emergent universe is suppressed at $l<30$.

Comparing with the results in Ref.~\cite{Huang2022}, we can find that the primordial power spectrum and CMB TT-spectrum in those models are the same as ours. However, the Einstein static universe in our model is stable against both scalar and tensor perturbations and the emergent universe can be realized successfully, whereas in the case of Ref.~\cite{Huang2022} unstable against inhomogeneous scalar perturbation.

\section{Conclusion}

In this paper, we have discussed the CMB power spectrum of the emergent universe with k-essence. We analyze the stability of the Einstein static universe against both scalar and tensor perturbations. When the inhomogeneous scalar perturbations are taken into consideration, the stable regions of the Einstein static universe are compressed. We find that the stable Einstein static universe can exist in the spatially closed spacetime and the stability conditions are given in Eq.~(\ref{stable}). To realize the universe exiting from the stable Einstein static state and evolving into an subsequent inflationart era, we construct a scalar potential ~(\ref{V2}) to break the stable condition of the Einstein static universe and assume a form of the equation of state parameter ~(\ref{w1}) to guarantee inflation occurs. An evolutionary curve of the scale factor $a$ is shown in Fig.~(\ref{Fig3}) which shows the emergent universe can be realized in this theory. Thus, the big bang singularity can be avoided in general relativity with k-essence.

As shown in Fig.~(\ref{Fig3}), since the numerical and analytical solution of the Friedmann equations overlap, the evolution of scale factor $a$ during inflation can be described by Eq.~(\ref{ai}) entirely. By considering the evolutionary form of the scale factor and solving the curved Mukhanov-Sasaki equation ~(\ref{MS}), we obtain the analytical approximation for the primordial power spectrum in the emergent universe. Then, we depict the primordial power spectrum and CMB TT-spectrum in Fig.~(\ref{Fig4}) which shows the primordial power spectrum is suppressed at $k<30$ and CMB TT-spectrum is suppressed at $l<30$.

\begin{acknowledgments}

This work was supported by the National Natural Science Foundation of China under Grants Nos. 11865018, 12265019, 11865019, 11505004, the regional first-class discipline of Guizhou province of China under Grants No. QJKYF[2018]216, the Doctoral Foundation of Zunyi Normal University of China under Grants No. BS[2017]07, and the Academic New Seedling Cultivation and Innovation Exploration Project of Zunyi Normal University of China under Grants No. ZunshiXM[2021]1-2.

\end{acknowledgments}


\begin{thebibliography}{99}

\bibitem{Guth1981} A. Guth, Phys. Rev. D {\bf 23}, 347 (1981).

\bibitem{Linde1982} A. Linde, Phys. Lett. {\bf 108B}, 389 (1982).

\bibitem{Albrecht1982} A. Albrecht and P. J. Steinhardt, Phys. Rev. Lett. {\bf 48}, 1220 (1982).

\bibitem{Ellis2004a} G. Ellis and R. Maartens, Class. Quantum Grav. {\bf 21}, 223 (2004).

\bibitem{Ellis2004b} G. Ellis, J. Murugan, and C. Tsagas, Class. Quantum Grav. {\bf 21}, 233 (2004).



\bibitem{Campo2007} S. Campo, R. Herrera, and P. Labrana, JCAP  {\bf 11}, 030 (2007).

\bibitem{Wu2010} P. Wu and H. Yu, Phys. Rev. D {\bf 81}, 103522 (2010).

\bibitem{Cai2012} Y. Cai, M. Li, and X. Zhang, Phys. Lett. B {\bf 718}, 248 (2012).

\bibitem{Zhang2014} K. Zhang, P. Wu, and H. Yu, JCAP  {\bf 01}, 048 (2014).

\bibitem{HuangQ2015} Q. Huang, P. Wu, and H. Yu, Phys. Rev. D {\bf 91}, 103502 (2015).

\bibitem{Shabani2017} H. Shabani and A. Ziaie, Eur. Phys. J. C {\bf 77}, 31 (2017).

\bibitem{Shabani2019} H. Shabani and A. Ziaie, Eur. Phys. J. C {\bf 79}, 270 (2019).

\bibitem{Huang2020} Q. Huang, B. Xu, H. Huang, F. Tu, and R. Zhang, Class. Quantum Grav. {\bf 37}, 195002 (2020).



\bibitem{Barrow2003} J. Barrow, G. Ellis, R. Maartens, and C. Tsagas, Class. Quantum Grav. {\bf 20}, L155 (2003).

\bibitem{Bohmer2015} C. Bohmer, N. Tamanini, and M. Wright, Phys. Rev. D {\bf 92}, 124067 (2015).

\bibitem{Huang2018a} Q. Huang, P. Wu, and H. Yu, Eur. Phys. J. C {\bf 78}, 51 (2018).

\bibitem{Huang2018b} Q. Huang, H. Huang, J. Chen, and S. Kang, Ann. Phys. {\bf 399}, 124 (2018).

\bibitem{Zhang2016} K. Zhang, P. Wu, H. Yu, and L. Luo, Phys. Lett. B {\bf 758}, 37 (2016).

\bibitem{Huang2014} H. Huang, P. Wu, and H. Yu, Phys. Rev. D {\bf 89}, 103521 (2014).

\bibitem{Li2017} S. Li and H. Wei, Phys. Rev. D {\bf 95}, 023531 (2017).

\bibitem{Bohmer2013} C. Bohmer, F. Lobo, and N. Tamanini, Phys. Rev. D {\bf 88}, 104019 (2013).

\bibitem{Atazadeh2017} K. Atazadeh and F. Darabi, Phys. Dark Universe {\bf 16}, 87 (2017).

\bibitem{Sharif2019} M. Sharif and A. Waseem, Astrophys. Space Sci. {\bf 364}, 221 (2019).

\bibitem{Sharif2018} M. Sharif and A. Waseem, Eur. Phys. J. Plus {\bf 133}, 160 (2018).

\bibitem{Li2019} S. Li, H. Lu, H. Wei, P. Wu, and H. Yu, Phys. Rev. D {\bf 99}, 104057 (2019).

\bibitem{Lewis2000} A. Lewis, A. Challinor, and A. Lasenby, Astrophys. J. {\bf 538}, 473 (2000).

\bibitem{Bernardeau2002} F. Bernardeau, S. Colombi, E. Gaztanaga, and R. Scoccimarro, Phys. Rep. {\bf 367}, 1 (2002).

\bibitem{Smoot1992} G. Smoot \textit{et al}., Astrophys. J. {\bf 396}, L1 (1992).

\bibitem{Planck2020} Planck Collaboration, A$\&$A {\bf 641}, A6 (2020).

\bibitem{Bonga2016} B. Bonga, B. Gupt, and N. Yokomizo, JCAP {\bf 10}, 031 (2016).

\bibitem{Handley2019} W. Handley, Phys. Rev. D {\bf 100}, 123517 (2019).

\bibitem{Thavanesan2021} A. Thavanesan, D. Werth, and W. Handley, Phys. Rev. D {\bf 103}, 023519 (2021).

\bibitem{Shumaylov2022} Z. Shumaylov and W. Handley, Phys. Rev. D {\bf 105}, 123532 (2022).

\bibitem{Dudas2012} E. Dudas, N. Kitazawa, S. P. Patil, and A. Sagnotti, JCAP {\bf 05}, 012 (2012).

\bibitem{Cai2015} Y. Cai, Y. Wang, and Y. Piao, Phys. Rev. D {\bf 92}, 023518 (2015).

\bibitem{Cai2018} Y. Cai, Y. Wang, J. Zhao, and Y. Piao, Phys. Rev. D {\bf 97}, 103535 (2018).

\bibitem{Ooba2018} J. Ooba, B. Ratra, and N. Sugiyama, The Astrophysical Journal {\bf 869}, 34 (2018).


\bibitem{Arya2018} R. Arya, A. Dasgupta, G. Goswami, J. Prasad, and R, Rangarajan, JCAP, {\bf 02}, 043 (2018).

\bibitem{Feng2003} B. Feng and X. Zhang, Phys. Lett. B, {\bf 570}, 145 (2003).

\bibitem{Kawasaki2003} M. Kawasaki and F. Takahashi, Phys. Lett. B, {\bf 570}, 151 (2003).


\bibitem{Labrana2015} P. Labrana, Phys. Rev. D {\bf 91}, 083534 (2015).

\bibitem{Huang2022} Q. Huang, K. Zhang, Z. Fang, and F. Tu, Phys. Dark Universe {\bf 38}, 101124 (2022).




\bibitem{Armendariz-Picon2000} C. Armendariz-Picon, V. Mukhanov, and P. Steinhardt, Phys. Rev. Lett. {\bf 85}, 4438 (2000).

\bibitem{Armendariz-Picon2001} C. Armendariz-Picon, V. Mukhanov, and P. Steinhardt, Phys. Rev. D {\bf 63}, 103510 (2001).

\bibitem{Armendariz-Picon1999} C. Armendariz-Picon, T. Damour, and V. Mukhanov, Phys. Lett. B {\bf 458}, 209 (1999).

\bibitem{Garriga1999} J. Garriga and V. F. Mukhanov, Phys. Lett. B {\bf 458}, 219 (1999).

\bibitem{Bose2009} N. Bose and A. S. Majumdar, Phys. Rev. D {\bf 79}, 103517 (2009).

\bibitem{Scherrer2004} R. Scherrer, Phys. Rev. Lett. {\bf 93}, 011301 (2004).

\bibitem{Aguirregabiria2004} J. Aguirregabiria, L. Chimento, and R. Lazkoz, Phys. Rev. D {\bf 70}, 023509 (2004).

\bibitem{Abramo2006} L. Abramo and N. Pinto-Neto, Phys. Rev. D {\bf 73}, 063522 (2006).

\bibitem{Yang2011} R. Yang and X. Gao, Class. Quantum Grav. {\bf 28}, 065012 (2011).

\bibitem{Chiba2009} T. Chiba, S. Dutta, and R. J. Scherrer, Phys. Rev. D {\bf 80}, 043517 (2009).

\bibitem{Bilic2008} N. Bilic, Phys. Rev. D {\bf 78}, 105012 (2008).

\bibitem{Barger2001} V. Barger and D. Marfatia, Phys. Lett. B {\bf 498}, 67 (2001).

\bibitem{Malquarti2003} M. Malquarti, E. Copeland, A. Liddle, and M. Trodden, Phys. Rev. D {\bf 67}, 123503 (2003).





\bibitem{Bardeen1980} J. M. Bardeen, Phys. Rev. D {\bf 22}, 1882 (1980).

\bibitem{Harrison1967} E. R. Harrison, Rev. Mod. Phys. {\bf 39}, 862 (1967).


\bibitem{Blas2011} D. Blas, J. Lesgourgues, and T. Tram, JCAP {\bf 07}, 034 (2011).



\end{thebibliography}
\end{document}